\documentclass[reprint,amsmath,amssymb,aps]{revtex4-1}

\usepackage{graphicx}
\usepackage{dcolumn}
\usepackage{bm}
\usepackage{hyperref}
\usepackage{epstopdf}
\usepackage{float}
\usepackage{color, soul}

\usepackage{pdfpages}

\begin{document}


\title{Winding light beams along elliptical helical trajectories}

\author{Yuanhui Wen$^1$, Yujie Chen$^{1,*}$, Yanfeng Zhang$^1$, Hui Chen$^1$, and Siyuan Yu$^{1,2,*}$}

\affiliation{$^1$State Key Laboratory of Optoelectronic Materials and Technologies, School of Electronics and Information Technology, School of Physics, Sun Yat-sen University, Guangzhou 510275, China
\\$^2$Photonics Group, Merchant Venturers School of Engineering, University of Bristol, Bristol BS8 1UB, UK
\\$^*$Corresponding authors: chenyj69@mail.sysu.edu.cn; s.yu@bristol.ac.uk}

\begin{abstract}
Conventional caustic methods in real or Fourier space produced accelerating optical beams only with convex trajectories. We develop a superposition caustic method capable of winding light beams along non-convex trajectories. We ascertain this method by constructing a one-dimensional (1D) accelerating beam moving along a sinusoidal trajectory, and subsequently extending to two-dimensional (2D) accelerating beams along arbitrarily elliptical helical trajectories. We experimentally implement the method with a compact and robust integrated optics approach by fabricating micro-optical structures on quartz glass plates to perform the spatial phase and amplitude modulation to the incident light, generating beam trajectories highly consistent with prediction. The theoretical and implementation methods can in principle be extended to the construction of accelerating beams with a wide variety of non-convex trajectories, thereby opening up a new route of manipulating light beams for fundamental research and practical applications.
\end{abstract}


\maketitle

The seemingly counter-intuitive discovery that light fields with appropriate initial field distributions such as Airy distribution \cite{PhysRevLett.99.213901,Siviloglou:07} capable of winding light beams as propagating along curved trajectories $-$ known as accelerating beams $-$ in free space has led to extensive research interests, as such optical phenomena may enable potential applications including optical micro-manipulation \cite{Nat.Photon.2008.201}, laser plasma filamentation \cite{Polynkin229}, 'light bullets' \cite{Siviloglou:07,chong2010airy,PhysRevLett.105.253901}, imaging \cite{Nat.Photon.2014.13}, and plasmonics \cite{Salandrino:10,Liu:11,PhysRevLett.107.116802,PhysRevLett.107.126804,PhysRevLett.109.093904}. Nevertheless, Airy beams only move along parabolic trajectories, which limit the flexibility in practical applications, so a number of researches have been carried out to extend the variety of beam trajectories. Direct phase engineering in real space under the paraxial approximation successfully constructed non-broadening beams moving along arbitrary convex trajectories at the cost of giving up propagation invariance \cite{PhysRevLett.106.213902}. Going beyond the paraxial condition in real space enabled large-angle bending of the trajectory\cite{Froehly:11}. These works can be categorized as the caustic method in real space. In addition to phase engineering in real space, similar works have also been carried out in Fourier space \cite{PhysRevA.88.043809,Hu:13,Bongiovanni2015Efficient}. However, only convex trajectories are realized based on these caustic methods because the first derivative of the trajectory is required to be single-valued. Limited types of non-convex trajectories are obtained by other approaches, such as by superposing some specific beams (e.g., circular beams) to construct periodic accelerating beams over an overall convex trajectory \cite{Hu:13,PhysRevLett.108.163901,GREENFIELD:12,Mathis:13} or by imposing a spiral-shape phase or amplitude modulation to incident beams \cite{Jarutis:09,2040-8986-12-12-124002,Matijosius:10,ADOM:ADOM201400315}. 

On the other hand, most of the experimental realization of wingding light beams has been based on phase-only spatial light modulators (SLM). In order to achieve high quality beams with low side-lobes, both amplitude and phase modulation would be required. Some schemes have been proposed to implement both phase and amplitude modulations in the generation of accelerating beams. One scheme is to encode both phase and amplitude modulation into a phase-only SLM \cite{Mathis:13}, which is more suitable for a binary amplitude modulation, otherwise it will introduce unwanted phase change and noise from other diffraction orders. Another scheme is to position a printed amplitude mask \cite{Hu:13} or a linear filter \cite{schley2014loss} close to the phase-only SLM to realize additional amplitude modulation, but accurate alignment between the amplitude and phase modulation elements is a major problem in such a configuration. Techniques of generating structured light beams have experienced a general evolution from bulk optical elements toward compact and robust integrated optical devices, exemplified in the case of vortex beam in Refs. \cite{Cai363,sun2015integrated}. Applying such integrated approaches to the generation of accelerating beams should also enable the engineering of the optical phase, amplitude and even polarization in a more accurate, compact and robust fashion \cite{doi:10.1021/nl402039y}. 

\begin{figure}
{\centerline{\includegraphics[width=8.5cm]{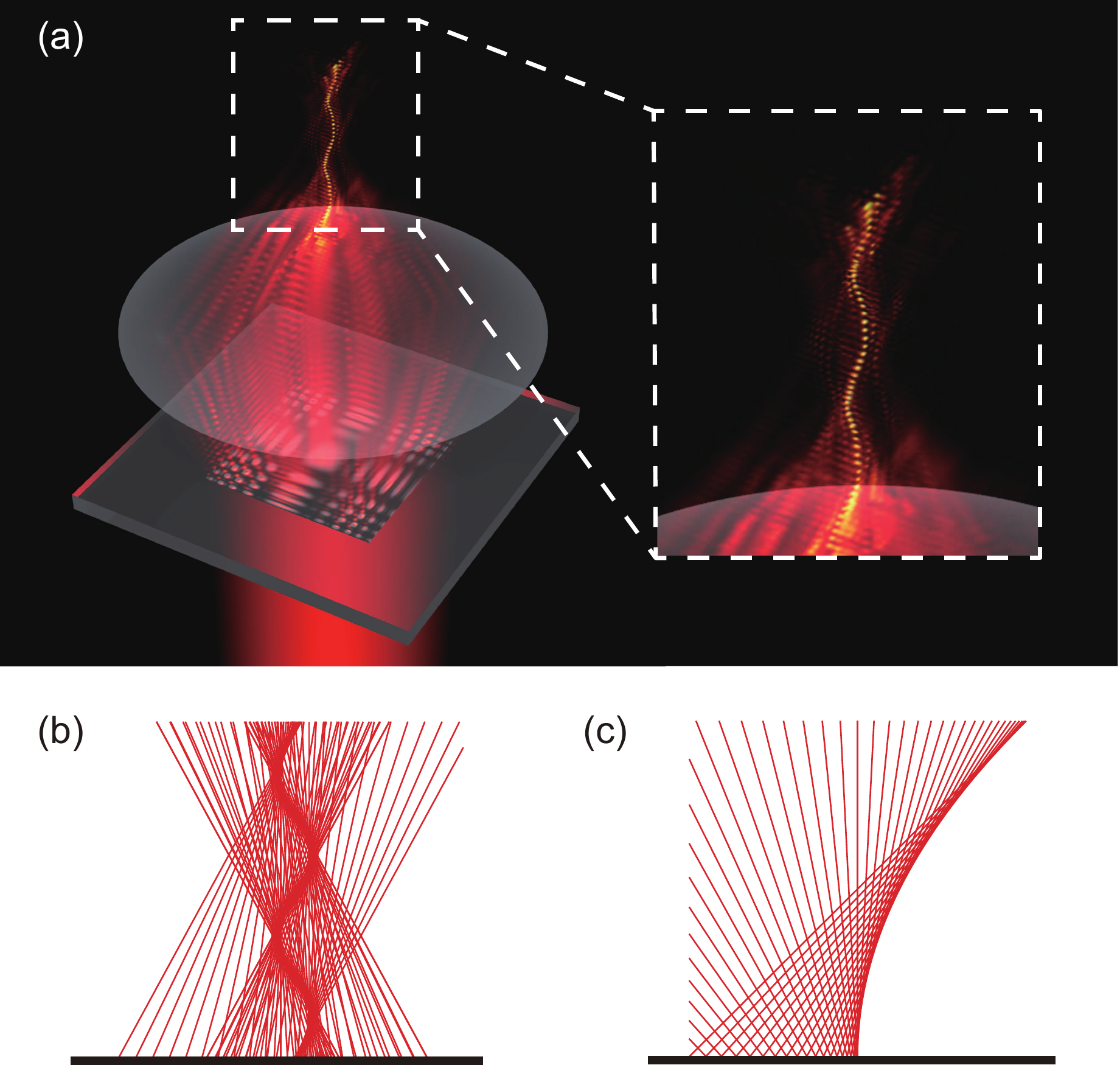}}}
\caption{\label{fig1}Schematic view and construction principle of an elliptical helical beam. (a) An illustration of the realization of a helical beam. (b) The transverse acceleration dynamics of 1D sinusoidal beam described as a bundle of rays compared with (c) the well-known Airy beam accelerating along a parabolic trajectory. The rays represent key spatial frequencies of light emanating from the initial plane (the thick black line at the bottom), and their interference forms a local intensity maximum along a curve.}
\end{figure}

In this work, we demonstrate the winding of light beams along arbitrarily elliptical helical trajectories in free space [Fig.~\ref{fig1}(a)] by developing a superposition caustic method and by means of an integrated optics approach.

A helical beam's 2D accelerating trajectory can be separated into two orthogonal 1D sinusoidal trajectories under the paraxial approximation. It is noted that the meaning of 1D and 2D here refers to the transverse profile of a light field and should not be confused with surface plasmonic waves propagating in a 2D plane \cite{Salandrino:10,Liu:11,PhysRevLett.107.116802,PhysRevLett.107.126804,PhysRevLett.109.093904}. Hence a 1D-accelerating beam with a sinusoidal trajectory is to be constructed first. Our method is based on the caustic method in Fourier space, where an accelerating beam moving along a curved trajectory is constructed by the interference of plane waves. Figures ~\ref{fig1}(b) and ~\ref{fig1}(c) present this interference process in a simplified light-ray picture with each red ray tangent to the predesigned trajectory representing the plane wave in need \cite{supplement}. The relation between a predesigned trajectory and its necessary plane waves in previous caustic method \cite{Hu:13} can be summarized as (see supplmental Material for a detailed derivation \cite{supplement})
\begin{equation}
\frac{{{k_x}}}{k} = f'(z), \theta ''({k_x}) = \frac{z}{k}, r({k_x}){\rm{ = }}{\left[ {\theta '''({k_x})/2} \right]^{1/3}}  
\label{eq:1}
\end{equation}
where $k_x$ and $k$ are the spatial frequency and wave number in free space, $r(k_x)$ and $\theta(k_x)$ are the amplitude and phase of the initial spatial spectrum $A(k_x)$, and $f(z)$ is the predesigned trajectory. Only if $f'(z)$ is single-valued can we obtain the necessary initial spatial spectrum, which limits the trajectory to be convex as mentioned before [Fig.~\ref{fig1}(c)]. 

To construct non-convex beams, we break this limitation by dividing the entire trajectory into several segments so that each segment is convex and thus corresponds to an initial spatial spectrum according to Eq.~\ref{eq:1}. If, over the space of one particular segment, the intensity along the main lobe of this segment is much higher than the side lobes in the same space resulting from the other segments, the interference between these segments along the main lobe is negligible, and the entire non-convex trajectory can be obtained by simple linear superposition of all the initial spatial spectra corresponding to all segments, which can be expressed as
\begin{equation}
A({k_x}) = \sum\limits_{j = 1}^n {{A_j}({k_x})}  = \sum\limits_{j = 1}^n {{r_j}({k_x}){e^{i{\theta _j}({k_x})}}}  
\label{eq:2}
\end{equation}
where the index $j$ indicates the ordinal of the segments along the propagation ($z-$) direction. In this manner, several positions on the trajectory are mapped onto the same spatial frequency, which is different from the one-to-one correspondence for convex trajectories constructed previously \cite{PhysRevLett.106.213902,Hu:13}. 

\begin{figure}[b]
{\centerline{\includegraphics[width=8.5cm]{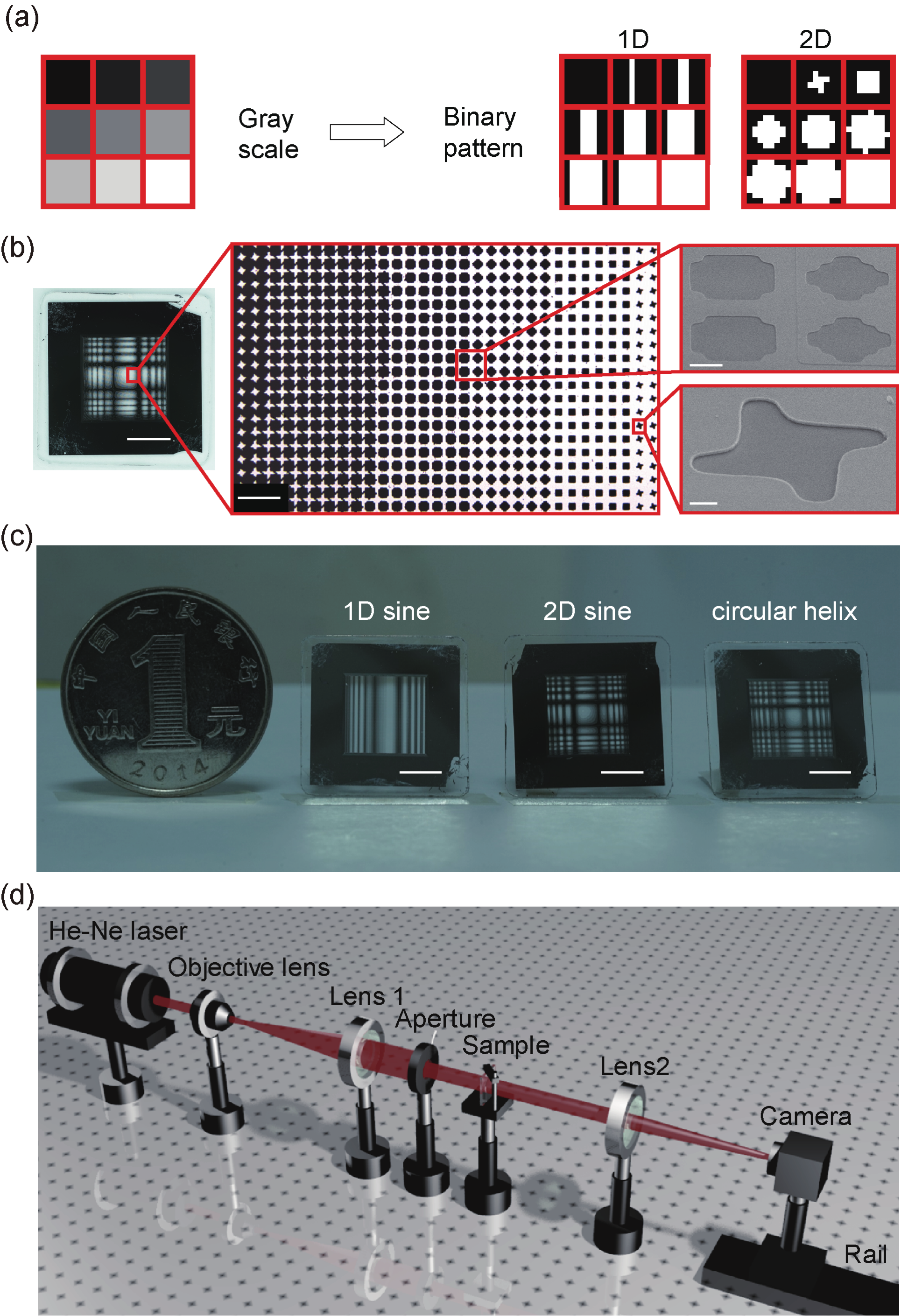}}}
\caption{\label{fig2}Experimental implementation. (a) 1D and 2D binary patterns employed to realize grey-scale amplitude modulation. (b) Enlarged view of the fabricated sample under the optical microscope and scanning electron microscope (SEM). The scale bar from left to right and from top to bottom is 5 mm, 50 $\mu$m, 5 $\mu$m, and 1 $\mu$m, respectively. (c) Three fabricated samples for generating the 1D, 2D sinusoidal beams and circular helical beam, respectively, with a scale bar of 5 mm. (d) Experimental setup used to generate the accelerating beams as well as measuring their propagation dynamics in free space.}
\end{figure}

\begin{figure*}
{\centerline{\includegraphics[width=14cm]{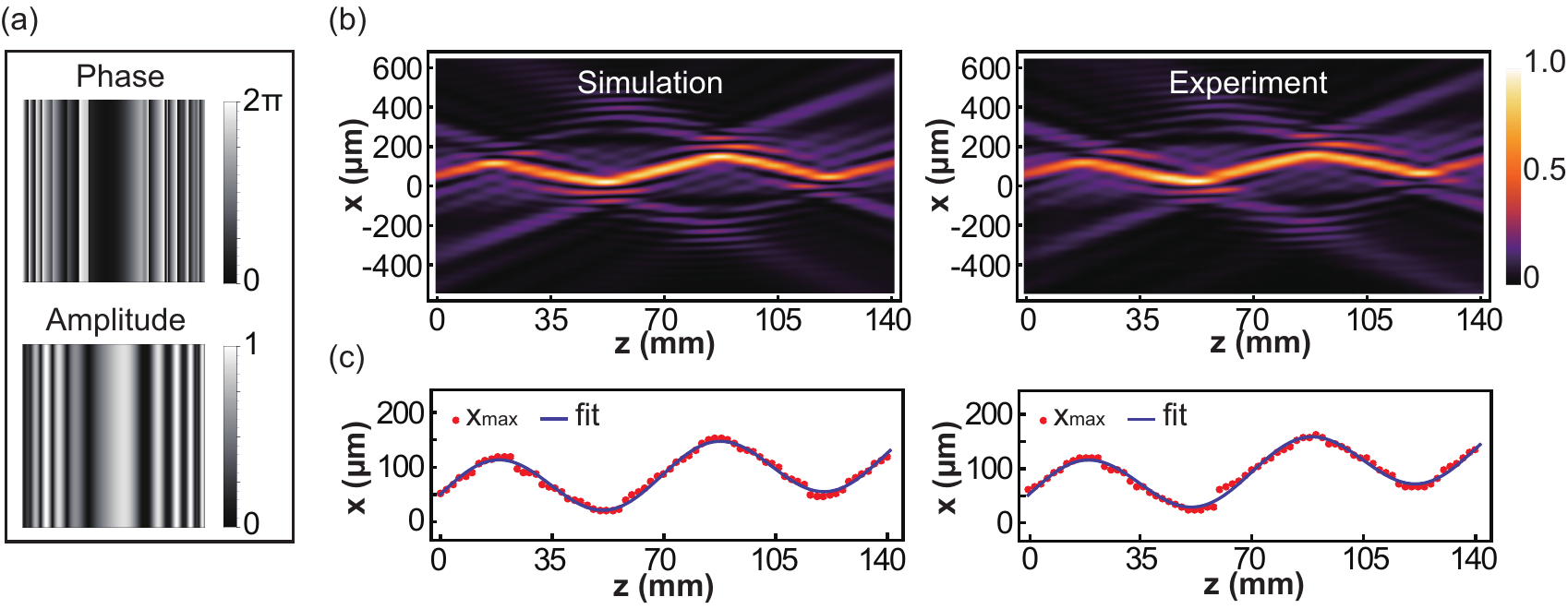}}}
\caption{\label{fig3}Generation of the basic 1D sinusoidal beam. (a) The designed phase and amplitude modulation. (b) Simulated and experimental intensity distribution for the beam propagating in free space. (c) Quantitative analysis of the main lobe trajectory. Here red dots mark the positions of the intensity maxima at the cross sections and the blue curves are the fitted trajectories. The fitted curve in simulation is $x = 50.6 + 0.49z + 54.2\sin \left( {2\pi z/69.3} \right)$, while in experiment is $x = 51.5 + 0.61z + 53.4\sin \left( {2\pi z/70} \right)$. The spatial unit is micron for $x$ and millimeter for $z$.}
\end{figure*}

Using this superposition caustic method, we are able to connect the beam trajectory with a specific initial phase and amplitude distribution. To perform the necessary phase and amplitude modulation to the incident light beam in experiment, integrated optics approach is exploited to fabricate a compact and robust optical element. The integrated optics element is fabricated on a quartz glass plate with a size of 20 mm $\times$ 20 mm $\times$ 0.5 mm and the effective area with patterns is 9.606 mm $\times$ 9.606 mm. This is divided into 601 $\times$ 601 pixels with each pixel to be 16 $\mu$m $\times$ 16 $\mu$m, representing sufficiently high sampling rates in the Fourier space. The phase modulation is implemented by aligned 16-level etching into the fused quartz plate. In each step, the structure is defined using photolithography in a positive photoresist (Shipley S1805), followed by a reactive ion etching process to transfer the pattern into the quartz plate, with each etch depth to be around 87 nm deep. The process is repeated for 15 times, overlaid using pre-fabricated alignment marks, so that the phase shift caused by each pixel is defined by the number of times it is etched. The amplitude modulation is realized by controlling the duty ratio of the partial metal coverage over the pixel. More specifically, each pixel is sub-divided into an $8 \times 8$ grid with each grid to be 2 $\mu$m $\times$ 2 $\mu$m, and the number of grids with metal coverage is controlled to realize specific transmittance for the whole pixel. Here nine different patterns are used to achieve a 9-level quasi-grey-scale [Fig.~\ref{fig2}(a)]. These patterns are designed with the appropriate symmetry in mind in order to minimize additional diffraction effects. The structure is patterned in the S1805 resist using photolithography. After development, an 80-nm Chromium layer was coated using electron beam evaporation and then followed by the standard lift-off process [Fig.~\ref{fig2}(b)]. The thickness of the whole micro-optical structures is less than 1.5 $\mu$m for the wavelength of 632.8 nm, with the alignment of the phase and amplitude modulation patterns to sub-micron precision, resulting in a highly compact and robust device for accelerating beam generation. Three fabricated elements for winding light beams are shown in Fig.~\ref{fig2}(c). 

After completion of the fabrication, an experimental setup, as shown in Fig.~\ref{fig2}(d), is used to generate the pre-designed accelerating beams as well as investigating their propagation dynamics in free space. A He-Ne laser operating at 632.8 nm emits a linearly polarized Gaussian beam that is expanded and collimated by an objective lens and lens 1. An aperture is used to obtain a quasi-plane wave before illuminating the fabricated quartz plate. After transmitting through the plate, the beam passes through lens 2 with a focal length $f$ = 500 mm, placed at a distance $f$ behind the quartz plate to perform the optical Fourier transform. The pre-designed accelerating beam is formed around a distance f behind lens 2, where a CMOS camera is placed to record the propagation dynamics of this beam by moving along the linear translating rail.

For the 1D sinusoidal beam, the designed initial phase and amplitude modulations varying along only one direction are shown in Fig.~\ref{fig3}(a). The beam propagation resulting from this modulation is numerically simulated with a Beam Propagation Method (OptiBPM) [Fig.~\ref{fig3}(b)]. A plot of the transverse position of intensity maxima along the z-direction shows that the trajectory fits very well with the sinusoidal curve, apart from a linear term representing a tilt of the beam axis [Fig.~\ref{fig3}(c)]. The experimentally measured beam pattern and trajectory are plotted [Fig.~\ref{fig3}(b) and ~\ref{fig3}(c)] and the trajectory coincides very well with the simulated trajectory, only the linear tilt is slightly different resulting from a slight tilt of the incident beam in the experiment. Moreover, the intensity variance along the main lobe shown in Supplementary \cite{supplement} Fig. S3 also has good agreement.

\begin{figure*}
{\centerline{\includegraphics[width=17.6cm]{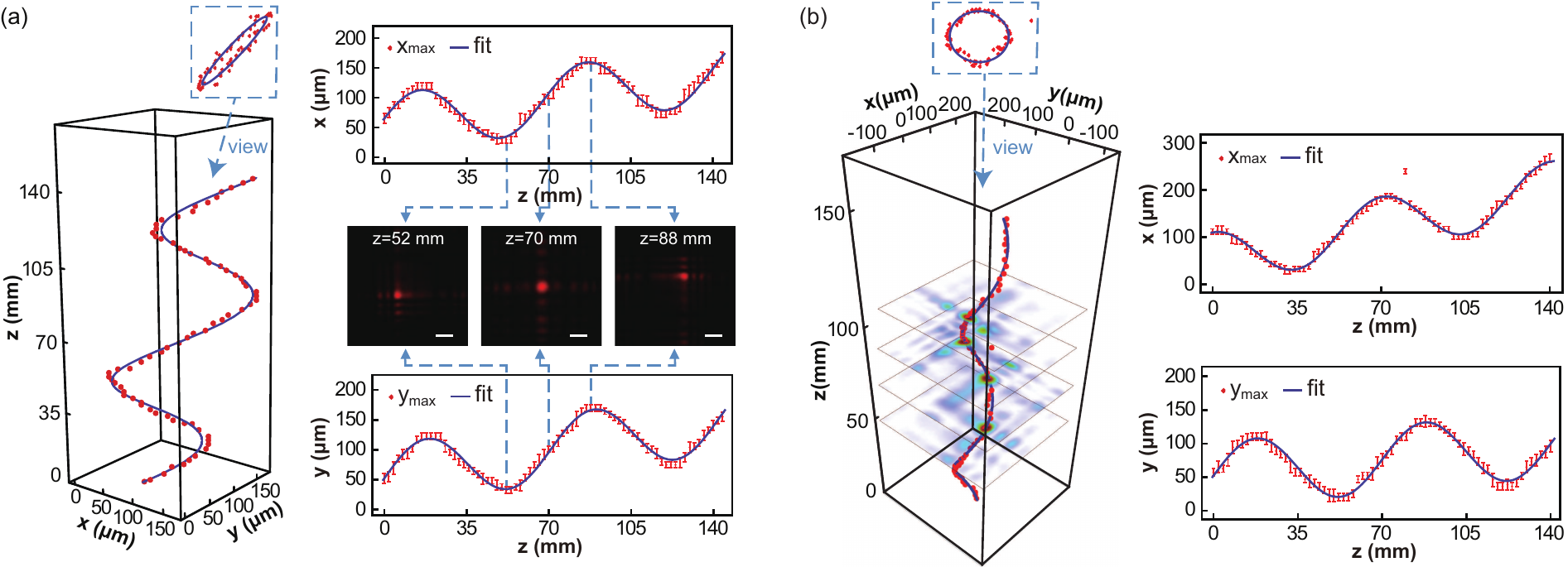}}}
\caption{\label{fig4}Realization of  a 2D sine beam and a helical beam based on the 1D sine beam. (a) The 3D plot of the trajectory for the main lobe of the 2D sine beam, and its projection to $x-z$ and $y-z$ planes (with error bar showing possible deviation in determining the position of the intensity maximum limited by our CMOS camera), as well as three selected cross sectional intensity distributions, with a scale bar of $100~\mu$m. Here the fitted curves in $x-z$ and $y-z$ planes are $x = 51.5 + 0.65z + 51.5\sin \left( {2\pi z/71{\rm{ + }}0.09\pi } \right)$ and $y = 51.5 + 0.69z + 54.3\sin \left( {2\pi z/71} \right)$ ($x$ and $y$ in micron, and $z$ in millimeter), respectively.  (b) The 3D illustration of the trajectory for the main lobe of the helical beam with four cross sectional intensity distributions presented and its projection to $x-z$ and $y-z$ planes. In this case, the fitted curves in $x-z$ and $y-z$ planes become $x = 51.5 + 1.08z + 57.6\sin \left( {2\pi z/70{\rm{ + }}\pi /2} \right)$ and $y = 51.5 + 0.35z + 49.6\sin \left( {2\pi z/70} \right)$ ($x$ and $y$ in micron, and $z$ in millimeter), respectively, with an introduced $\pi$/2 phase shift as designed.}
\end{figure*}

Due to the variables of $x$ and $y$ being separable under the paraxial condition, it is straightforward to construct 2D accelerating beams simply by multiplying two 1D field distributions along $x$ and $y$ direction \cite{Siviloglou:07,Bongiovanni2015Efficient}. Therefore, 2D accelerating beams here with arbitrarily elliptical helical trajectories can be constructed based on two 1D sinusoidal trajectories in $x-z$ and $y-z$ planes respectively. If the two 1D sinusoids are in phase, the helical beam degenerates into a 2D sinusoidal beam, whose projection on the $x-y$ plane is a line along the ${45^ \circ}$ angle. The experimental result is shown in Fig.~\ref{fig4}(a) and its projections to the $x-z$ and $y-z$ planes are similar to the 1D sine beam. Although its projection onto the transverse plane along the ${47^ \circ}$ angle broadens slightly to an ellipsoid indicating a small phase difference of $\sim 0.09\pi$ which may result from a slight tilt of the spherical lenses in the experimental setup leading to unequal focal lengths, the experimental results are still highly consistent with the simulation (Supplementary \cite{supplement} Fig. S4).

Furthermore, if a phase shift $\delta$ is introduced between the trajectories in the $x-z$ and $y-z$ planes, winding light beams along arbitrarily elliptical helical trajectories can be realized with the eccentricity $\sqrt {{{2\left| {\cos \delta } \right|} \mathord{\left/
 {\vphantom {{2\left| {\cos \delta } \right|} {\left( {1 + \left| {\cos \delta } \right|} \right)}}} \right.
 \kern-\nulldelimiterspace} {\left( {1 + \left| {\cos \delta } \right|} \right)}}} $ ranging from $0$ to $1$. For demonstration, we choose $\delta$ to be $\pi /2$ and thus the eccentricity becomes zero, corresponding to a circular helical beam. The necessary phase and amplitude modulation as well as the experimental result are shown in Fig.~\ref{fig4}. In this case, except one obvious point deviating from the fitted curve resulting from the imperfect fabrication of the quartz glass plate (detailed analysis in Supplementary \cite{supplement} Fig. S5), the trajectory also coincides well with the simulation results (Supplementary \cite{supplement} Fig. S6).

In conclusion, by way of a superposition caustic method, we break the one-to-one correspondence between the real space position on the beam trajectory and the spatial frequency in the Fourier space caustic method, which enables the construction of accelerating beams whose main lobe moves along non-convex sinusoidal trajectories. We experimentally realize 1D sinusoidal accelerating beams using this method. We further construct 2D accelerating beams with arbitrarily elliptical helical trajectories by superimposing two 1D sinusoidal trajectories in the $x-z$ and $y-z$ planes and controlling their phase shift. In principle, such a method could be applied to the construction of accelerating beams along a wide variety of other non-convex trajectories.

We employ an integrated optics approach to the experimental implementation of winding light beams by fabricating micro-optical structures on quartz glass plates to simultaneously provide the designed phase and amplitude modulation with high precision. The method is compact and robust, as confirmed by the high consistency between the experiment and the simulation results, and paves the way for the reliable generation of various accelerating beams by means of photonic integration for practical applications in the future. 

$\\$
This work is supported by the National Basic Research Program of China (973 Program) (2014CB340000 and 2012CB315702), the Natural Science Foundations of China (61323001, 61490715, 51403244, and 11304401), the Natural Science Foundation of Guangdong Province (2014A030313104), and the Fundamental Research Funds for the Central Universities of China (Sun Yat-sen University: 13lpgy65, 15lgpy04, 15lgzs095, 15lgjc25, and 16lgjc16). Y. Chen would also like to thank the Specialized Research Fund for the Doctoral Program of Higher Education of China (20130171120012). Y. Wen thanks the Undergraduate Training Programs for Innovation and Entrepreneurship (201302071) and SPE's Undergraduate Scientific Research Program (2013011). The authors thank Lin Liu, Chunchuan Yang and Zengkai Shao for technical assistance in micro-fabrication, as well as Pengfei Xu, Jiangbo Zhu, Yi Wang, Guoxuan Zhu, and Johannes Herrnsdorf (University of Strathclyde, UK) for helpful discussions.


%

\newpage 
$ $
\newpage 

\includepdfmerge{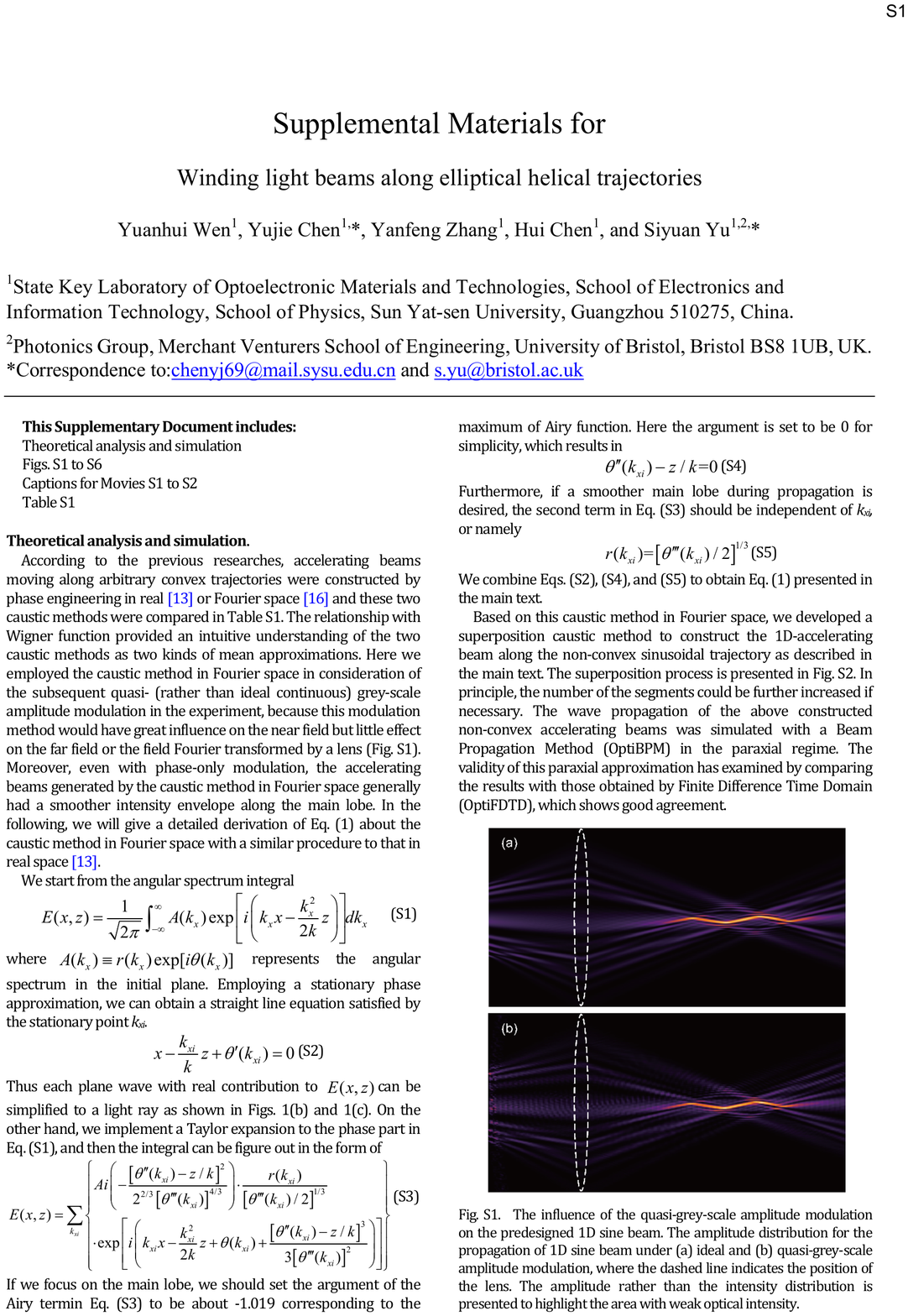,1}

\newpage 
$ $
\newpage 
\includepdfmerge{Supplemental_Materials.pdf,2}

\newpage 
$ $
\newpage 


\section*{Supplementary material}

# Supplemental Materials for

## Winding light beams along elliptical helical trajectories

Yuanhui Wen[1], Yujie Chen[1,*], Yanfeng Zhang[1], Hui Chen[1], and Siyuan Yu[1,2,*]

[1]State Key Laboratory of Optoelectronic Materials and Technologies, School of Electronics and Information Technology, School of Physics, Sun Yat-sen University, Guangzhou 510275, China.

[2]Photonics Group, Merchant Venturers School of Engineering, University of Bristol, Bristol BS8 1UB, UK.
*Correspondence to: chenyj69@mail.sysu.edu.cn and s.yu@bristol.ac.uk

**This Supplementary Document includes:**
Theoretical analysis and simulation
Figs. S1 to S6
Captions for Movies S1 to S2
Table S1

**Theoretical analysis and simulation.**

According to the previous researches, accelerating beams moving along arbitrary convex trajectories were constructed by phase engineering in real [13] or Fourier space [16] and these two caustic methods were compared in Table S1. The relationship with Wigner function provided an intuitive understanding of the two caustic methods as two kinds of mean approximations. Here we employed the caustic method in Fourier space in consideration of the subsequent quasi- (rather than ideal continuous) grey-scale amplitude modulation in the experiment, because this modulation method would have great influence on the near field but little effect on the far field or the field Fourier transformed by a lens (Fig. S1). Moreover, even with phase-only modulation, the accelerating beams generated by the caustic method in Fourier space generally had a smoother intensity envelope along the main lobe. In the following, we will give a detailed derivation of Eq. (1) about the caustic method in Fourier space with a similar procedure to that in real space [13].

We start from the angular spectrum integral

$$E(x,z) = \frac{1}{\sqrt{2\pi}} \int_{-\infty}^{\infty} A(k_x) \exp\left[i\left(k_x x - \frac{k_x^2}{2k} z\right)\right] dk_x \quad (S1)$$

where $A(k_x) \equiv r(k_x)\exp[i\theta(k_x)]$ represents the angular spectrum in the initial plane. Employing a stationary phase approximation, we can obtain a straight line equation satisfied by the stationary point $k_{xi}$.

$$x - \frac{k_{xi}}{k} z + \theta'(k_{xi}) = 0 \quad (S2)$$

Thus each plane wave with real contribution to $E(x,z)$ can be simplified to a light ray as shown in Figs. 1(b) and 1(c). On the other hand, we implement a Taylor expansion to the phase part in Eq. (S1), and then the integral can be figure out in the form of

$$E(x,z) = \sum_{k_{xi}} \left\{ Ai\left(-\frac{[\theta''(k_{xi}) - z/k]^2}{2^{2/3}[\theta'''(k_{xi})]^{4/3}}\right) \cdot \frac{r(k_{xi})}{[\theta'''(k_{xi})/2]^{1/3}} \cdot \exp\left[i\left(k_{xi} x - \frac{k_{xi}^2}{2k} z + \theta(k_{xi}) + \frac{[\theta''(k_{xi}) - z/k]^3}{3[\theta'''(k_{xi})]^2}\right)\right] \right\} \quad (S3)$$

If we focus on the main lobe, we should set the argument of the Airy term in Eq. (S3) to be about -1.019 corresponding to the maximum of Airy function. Here the argument is set to be 0 for simplicity, which results in

$$\theta''(k_{xi}) - z/k = 0 \quad (S4)$$

Furthermore, if a smoother main lobe during propagation is desired, the second term in Eq. (S3) should be independent of $k_{xi}$, or namely

$$r(k_{xi}) = [\theta'''(k_{xi})/2]^{1/3} \quad (S5)$$

We combine Eqs. (S2), (S4), and (S5) to obtain Eq. (1) presented in the main text.

Based on this caustic method in Fourier space, we developed a superposition caustic method to construct the 1D-accelerating beam along the non-convex sinusoidal trajectory as described in the main text. The superposition process is presented in Fig. S2. In principle, the number of the segments could be further increased if necessary. The wave propagation of the above constructed non-convex accelerating beams was simulated with a Beam Propagation Method (OptiBPM) in the paraxial regime. The validity of this paraxial approximation has examined by comparing the results with those obtained by Finite Difference Time Domain (OptiFDTD), which shows good agreement.

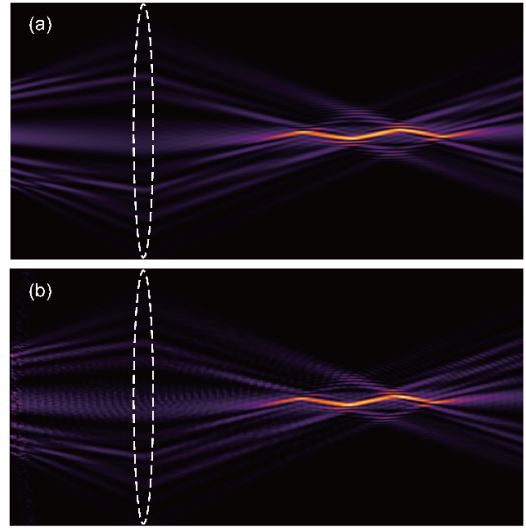

Fig. S1. The influence of the quasi-grey-scale amplitude modulation on the predesigned 1D sine beam. The amplitude distribution for the propagation of 1D sine beam under (a) ideal and (b) quasi-grey-scale amplitude modulation, where the dashed line indicates the position of the lens. The amplitude rather than the intensity distribution is presented to highlight the area with weak optical intensity.

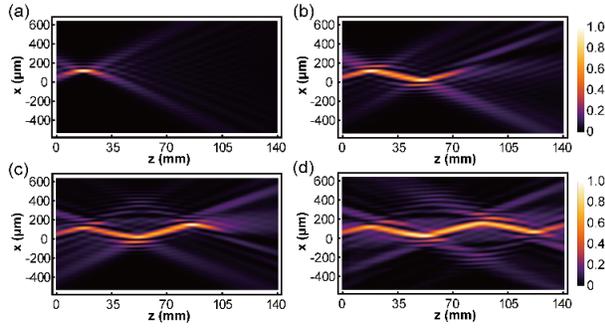

Fig. S2. The intensity distribution of the 1D sine beams with increasing number of segments in the superposition process. The 1D sine beam with two periods is selected to demonstrate in the main text, but it is worth noting that more bends are available in principle.

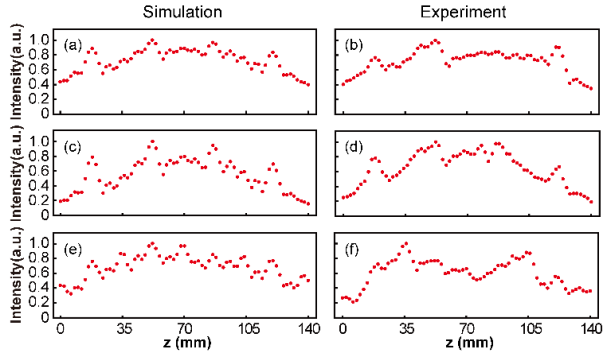

Fig. S3. Comparison of the Intensity variation along the main lobe in simulation and experiment. The intensity distribution along the main lobe of the (a), (b) 1D sine beams, (c), (d) 2D sine beams and (e), (f) helical beams in simulation and experiment are presented. As shown clearly, the simulation and experiment results are consistent on the whole, especially for the 1D and 2D sine beams, and the deviation in detail results from the quasi-grey-scale amplitude modulation rather than an ideal amplitude modulation.

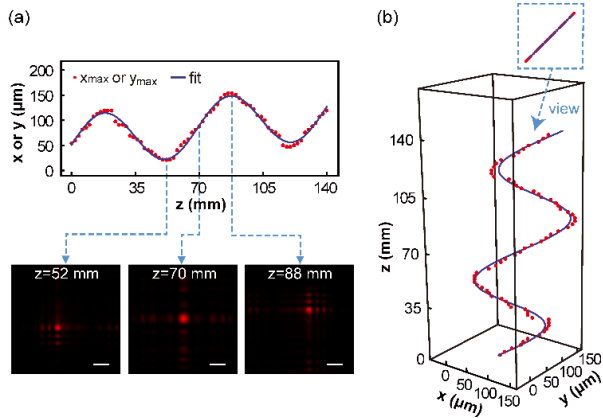

Fig. S4. Simulation results of the 2D sine beam. (a) The trajectories of the main lobe projected to the *x-z* and *y-z* planes are designed to be the same (in phase) with the fitted curve to be $x$ or $y = 51.5 + 0.49z + 54.3\sin(2\pi z/69.3)$ (the unit is micron for *x, y* and millimeter for *z*). And three selected cross-sectional intensity distributions corresponding to the experiment in figure 3 are presented, with a scale bar of 100 μm. (b) The 3D illustration of the trajectory.

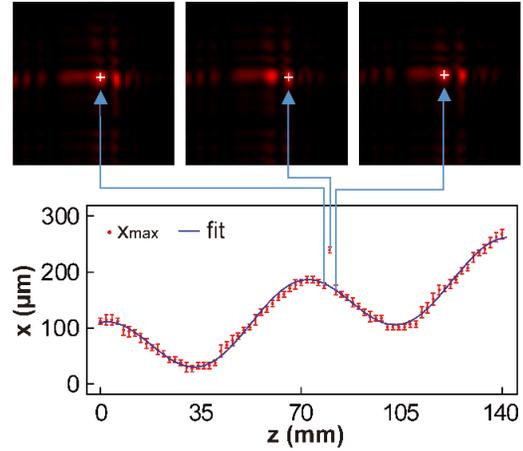

Fig. S5. Intensity distribution of the three cross sections around the deviation point in Fig. 4(b). The experimental result of the circular helical beam [Fig. 4(b)] has a main lobe maximum point obviously deviating from the fitted curve in the *x* direction. Here we have shown the intensity distribution of three cross sections around that point, and we find that the deviation reveals the jump of the intensity maximum (marked by a cross) from the original main lobe to the side lobe in that cross section, which may result from the imperfect metal evaporation in fabricating the quartz glass plate and thus an unsatisfactory amplitude modulation.

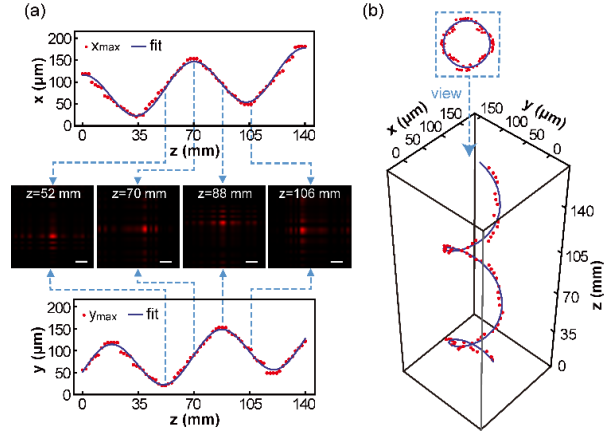

Fig. S6. Simulation results of the circular helical beam. (a) The trajectories of the main lobe projected to the *x-z* and *y-z* planes are designed to have a π/2 phase shift between them. The fitted curves along *x-z* and *y-z* planes are $x = 61.8 + 0.44z + 53.6\sin(2\pi z/69.3 + \pi/2)$ and $y = 51.5 + 0.49z + 54.2\sin(2\pi z/69.3)$ (the unit is micron for *x, y* and millimeter for *z*). Four selected cross-sectional intensity distributions corresponding to the experiment are also presented, with a scale bar of 100 μm. (b) The 3D illustration of the trajectory.

Movie S1

The movie shows the 3D reconstruction of the propagation dynamics of the 2D sine beam in simulation and experiment. The propagation process is shown first, where the main lobe trajectory is not continuous because of finite discretized cross-sectional data used (in order to be better view its bending). A rendition of both beams with changing perspective around their own axis is then used to further reveal the 3D beam trajectory.

Movie S2

The movie shows the 3D reconstruction of the propagation dynamics of the circular helical beam in simulation and experiment. The propagation process is shown first, where the main lobe trajectory is not continuous because of finite discretized cross-sectional data used (in order to be better view its bending). A rendition of both beams with changing perspective around their own axis is then used to further reveal the 3D beam trajectory.

Table S1. Comparison of the caustic methods in real and Fourier space.

| | Caustic method in Fourier space | Caustic method in real space |
|---|---|---|
| Diffraction integral | $E(x,z) = \frac{1}{\sqrt{2\pi}} \int_{-\infty}^{\infty} A(k_x) e^{i\left[k_x x + (k - \frac{k_x^2}{2k})z\right]} dk_x$ where $A(k_x) \equiv r(k_x) e^{i\theta(k_x)}$ | $E(x,z) = \frac{e^{ikz}}{\sqrt{i\lambda z}} \int_{-\infty}^{\infty} E(x_0, 0) e^{\frac{ik}{2z}(x-x_0)^2} dx_0$ where $E(x_0, 0) \equiv \rho(x_0) e^{i\varphi(x_0)}$ |
| Light ray equation | $x = -\theta'(k_x) + \frac{k_x}{k} z$ | $x = x_0 + \frac{\varphi'(x_0)}{k} z$ |
| Derived results | $\begin{cases} \frac{k_x}{k} = f'(z), \quad \theta''(k_x) = \frac{z}{k}, \\ r(k_x) = [\theta'''(k_x)/2]^{1/3} \end{cases}$ | $\begin{cases} x_0 = f(z) - zf'(z) \\ \varphi''(x_0) = -\frac{1}{z} \end{cases}$ |
| Relationship with Wigner function | $\frac{\int_{-\infty}^{\infty} W(x, k_x, 0) x \, dx}{\int_{-\infty}^{\infty} W(x, k_x, 0) \, dx} = -\theta'(k_x)$ | $\frac{\int_{-\infty}^{\infty} W(x, k_x, 0) k_x \, dk_x}{\int_{-\infty}^{\infty} W(x, k_x, 0) \, dk_x} = \varphi'(x)$ |



\end{document}